\documentclass[
    ,final            
  ]
  {aipproc}

\layoutstyle{6x9}

\newcommand{\mpi}{m_\pi}

\newcommand{\gev}{\,{\rm GeV}}
\newcommand{\mev}{\,{\rm MeV}}

\begin{document}

\title{Unravelling strange quarks in nucleon structure}


\author{R.~D.~Young}{
  address={Special Research Centre for the Subatomic Structure of Matter (CSSM),\\
           School of Chemistry and Physics, University of Adelaide, SA 5005, Australia}
}


\begin{abstract}
  We present a discourse on the stages of discovery that have led to a
  deeper understanding of the role played by strange quarks in the
  structure of the nucleon.
\end{abstract}

\maketitle


\section{Introduction}
The determination of the strange quark content of the nucleon offers a
unique probe to measure the nonperturbative structure of the
nucleon. As the nucleon carries zero net strangeness, the influence of
strange quarks arises entirely through interaction with the vacuum. In
essence, the strange quarks play a role analogous to the Lamb shift in
QED.  While strangeness contributions to nucleon structure have been
difficult to isolate, the measurement of the neutral weak current in
elastic scattering offers perhaps the most direct observation of the
strange quark content of the nucleon \cite{PVES}. Here we highlight
advances in the theoretical determination of the strangeness
electromagnetic form factors, and compare with experimental
measurements. Further, we discuss recent work that has provided an
accurate determination of the strangeness sigma term based on a chiral
extrapolation of lattice QCD results.

\section{Chiral phenomenology of disconnected quarks}
In early lattice QCD simulations, extrapolations to the physical quark
masses largely neglected the importance of incorporating the dynamical
consequences of chiral symmetry breaking in QCD. Indeed it is evident
that (at least) part of the rationale for neglecting these features
was the empirical observation that lattice results displayed smooth
and slowly-varying dependence on the quark mass --- contrasting the
rapid nonlinear effects that must exist provided QCD's chiral symmetry
is spontaneously broken.  A solution to explain the rather linear
behaviour of the lattice results {\it and} incorporate the correct
dynamical constraints of QCD was identified in early work, see
Refs.~\cite{Leinweber:1999ig,Young:2002ib,Cloet:2003jm} for
instance. This work used a momentum-space cutoff (or finite-range
regularisation) to supress the high-momentum interations between
baryons and pions, such as to supress the interactions of pion-loop
dressings once the pion Compton wavelength is small relative to the
(axial) size of the baryon.

The success of this work was extended to baryon masses in quenched
lattice simulations \cite{Young:2002cj} --- which acted as a further
testing ground, given the more readily available quenched results at
the time. The chiral extrapolation of this work identified quenching
artifacts to be much larger than previously claimed \cite{Aoki:1999yr}
--- results that have since been confirmed by improved quenched
simulations in the chiral regime
\cite{Gattringer:2003qx,Zanotti:2004dr}.

A surprising result of the work in Ref.~\cite{Young:2002cj} was the
observation that the differences in quenched and dynamical baryon
masses\footnote{The quenched and dynamical simulations are ``matched''
  by choosing an intermediate-range (non-chiral) scale, such as $r_0$,
  to set the lattice spacing.} could be largely described by
finite-range regularised chiral loops. This discovery opened the
possibility of estimating the effects of quenching, and thereby
providing improved estimates of QCD observables from quenched
simulations.

This description was extended to the proton magnetic moment
\cite{Young:2004tb}, which enabled a determination of the proton
magnetic moment from quenched lattice simulations. To this date, the
{\em proton} magnetic moment has still not been calculated in
dynamical simulations --- as the disconnected contribution continues
to be neglected.

With a description of the unquenching effects in baryon properties,
which appears to work for masses and magnetic moments, one had the 
confidence to extend to more ambitious observables --- such as the
flavour separation of the nucleon electromagnetic form factors. In
particular, the isolation of the strangeness component of these form
factors.

In combination with charge-symmetry relations among the octet-baryon
magnetic moments \cite{Leinweber:1999nf}, our analysis enabled a
precise determination of the strangeness magnetic moment,
$G_M^s=-0.046\pm0.022\,\mu_N$ \cite{Leinweber:2004tc,gmserr}. A strong
indicator of the reliability of this analysis was the excellent
agreement found with the experimentally determined baryon magnetic
moments.

Further extensions within the same framework enabled the
determinations of the strangeness charge radius
\cite{Leinweber:2006ug} and the strangeness magnetic form factor at
$Q^2\sim 0.23\gev^2$ \cite{Wang:2008ta}. These results, especially the
magnetic moment, challenged the best experimental determinations at
the time.

\section{Strangeness measurements}
The strangeness contributions to the vector form factors of the
nucleon can be probed in parity-violation measurements. Early
combinations of different measurements suggested discrepancy with the
described theory calculation
\cite{Maas:2004dh,Aniol:2005zg,Armstrong:2005hs}, with $G_M^s$
indicated to be positive at roughly 2-sigma --- constrasting the
precise negative value shown above.

\begin{figure}
\resizebox{0.8\textwidth}{!}{\includegraphics{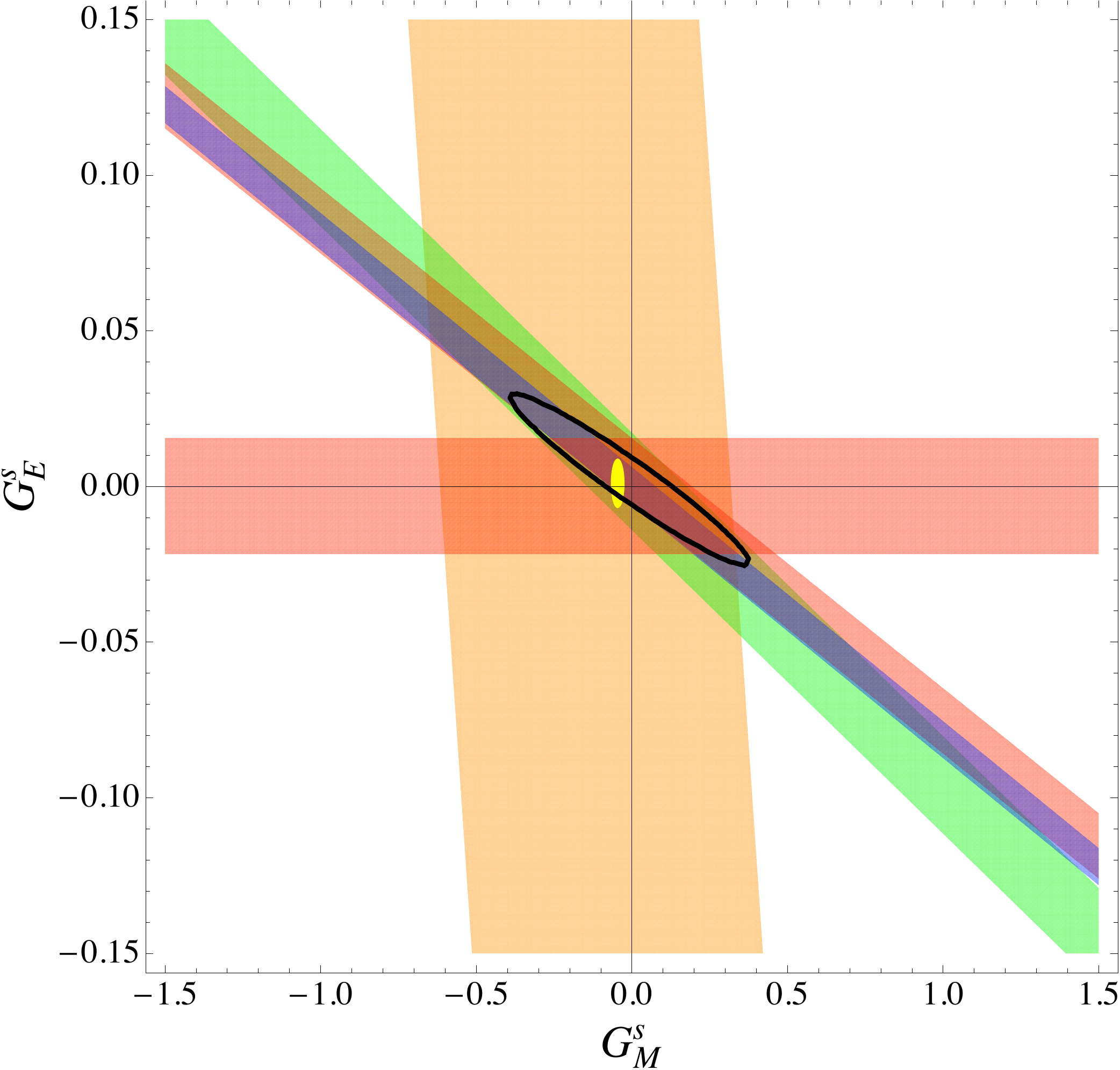}}
\caption{Summary of experimental determination of strangeness form
  factors at $Q^2=0.1\gev^2$. Open ellipse depicts the combined 68\%
  confidence interval of the experimental results. The filled ellipse
  displays the theory calculation described.}
\label{fig:strange}
\end{figure}

A comprehensive global analysis of the original experimental
asymmetries \cite{Young:2006jc} --- including a consistent treatment
of electromagnetic form factors and radiative corrections; a Taylor
expansion of $G_E^s$ and $G_M^s$; and an experimental extraction of
the anapole form factors --- led to an extraction of strangeness found
to be in much better agreement with the predictions outlined above.
Shortly after, this analysis was further supported by the
high-precision HAPPEX measurement \cite{Acha:2006my} on both hydrogen
and helium targets. A combined global analysis of the form factors at
that point in time is shown in Figure~\ref{fig:strange}.

Though yet to be incorporated in the global analysis, recent
back-angle results at slightly higher $Q^2$ also support the findings
of a small strangeness \cite{Baunack:2009gy,Androic:2009zu}.  A
recent direct lattice QCD calculation of the strangeness form
factors \cite{Doi:2009sq} also supports this scenario.

In addition to resolving the strange quarks in the nucleon, the
collective precision of the parity-violation measurements provides a
robust test of the electroweak interaction. The broad kinematic
coverage of G0 \cite{Armstrong:2005hs} and precision of HAPPEX
\cite{Acha:2006my} provides a reliable extrapolation to extract the
weak charge of the proton \cite{Young:2007zs}. In combination with
earlier atomic parity-violation measurements \cite{Bennett:1999pd},
the proton measurement places tight constraints on new electroweak
physics up to a characteristic energy scale of $\sim 1\,{\rm TeV}$.

\section{Strangeness sigma term}
While the unquenching component of the theoretical strangeness
analysis relied on the phenomenological description of sea quark
effects, the strength of the extrapolation to the physical point
relied heavily on the methods of finite-range regularization detailed
in Refs.~\cite{Young:2002ib,Leinweber:2003dg,Leinweber:2005xz}. Here
it was established that there is very minimal dependence on the choice
of regulator. With the latest generation of lattice simulations now
taking full account of the 3 light flavours of dynamical quarks, the
methods of FRR can now be utilised without the need to incorporate the
model-dependent unquenching component.

Recent lattice QCD results for the octet baryon masses, using
2+1-flavours of dynamical quarks \cite{WalkerLoud:2008bp,Aoki:2008sm},
have been extrapolated using an SU(3) chiral extrapolation with FRR
\cite{Young:2009zb}. The results of this analysis produce 
precise values for both the absolute baryon masses and the associated
sigma terms.

\begin{figure}
\includegraphics[width=0.8\columnwidth,angle=0]{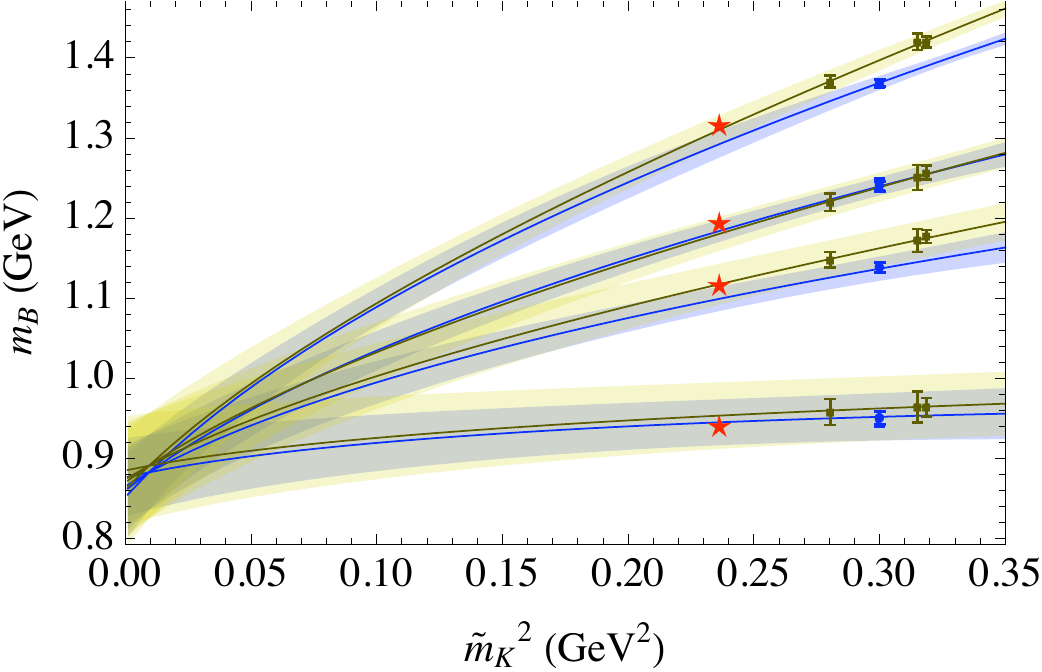}
\caption{Extapolation in the strange quark mass, holding the light
  quarks {\em fixed} at the physical point. The square data points and
  matching contours display the PACS-CS as extrapolated to the
  physical pion mass, while the circles are those of the LHPC
  extrapolation. From bottom--top displays the $N$, $\Lambda$,
  $\Sigma$ and $\Xi$ baryon, with the physical masses indicated by the
  stars.}
\label{fig:extrap}
\end{figure}

The full functional form of the fits permits a comprehensive
description of the baryon masses over a range of quark masses. In
Figure~\ref{fig:extrap} we show the dependence on the strange quark
mass, as it is taken from the lattice results and down through the
physical point to the strangeness chiral limit. This figure is plotted
against the SU(2) chiral limit kaon mass,
$\tilde{m}_K^2=m_K^2-\frac12\mpi^2$, which we use as an effective
measure of the deviation of the strange quark mass from the chiral
limit. The figure is shown for the pion mass being held fixed at the
physical point.
The lattice results are individually extrapolated to the physical pion
mass using the respective (LHPC or PACS-CS) fit result. The figure
indicates the effective spread in the strange-quark mass probed by
PACS-CS (the two LHPC points considered are at
essentially the same strange-quark mass). The error bars on the points
just show the original lattice error bar (including finite-volume
uncertainty), whereas the shaded region depicts the full
uncertaity at these points including the extrapolation to the physical
pion mass.

An important feature of this figure is the very weak dependence of the
nucleon mass on the strange quark. This leads to a particularly small value
for the strangeness sigma term, $\sigma_{Ns}=31\pm 15\mev$
\cite{Young:2009zb}. This value is compatible with an independent
direct lattice extraction by Toussaint and Freeman, $\sigma_{Ns}=59\pm
10\mev$ \cite{Toussaint:2009pz}.

We note that this consensus on a small strangeness sigma term is also
being supported by a 2-flavour lattice analysis of Ohki~{\it et
  al.}~\cite{Ohki:2008ff}, and preliminary results of their extensions
to the 3-flavour case \cite{Ohki:2009mt}.  Further, a recent chiral
analysis of the PACS-CS results above, using an alternative
regularization scheme, finds a compatible set of sigma terms
for the octet baryon ensemble \cite{MartinCamalich:2010fp} to our
determination report in Ref.~\cite{Young:2009zb}.

One consequence of this improved precision in the determination of the
strange quark sigma term is a dramatic reduction in the
uncertainties of dark matter cross sections in a range of
supersymmetric models \cite{Giedt:2009mr}. The predicted cross
sections are found to be substantially smaller than previously
suggested. While this is somewhat unfortunate from an experimental
perspective, the new level of precision does indicate that any
observation of dark matter would have substantial discrimination power
amongst the class of benchmark models that have been considered.

\section{Closing remarks}
It is evident that the strange quark contributions to the structure of
the nucleon are smaller than early estimates had suggested. Indeed, at
the 2-sigma level, the strangeness magnetic moment and mean-square
charge radius of the proton are less than about 6\% of their total
values. Conservatively, the strangeness sigma term appears to lie
somewhere in the range of just 2--7\% of the nucleon mass.

We look forward to the next decade of discoveries that will continue
to improve our understanding of the nonperturbative nucleon.

\section{Acknowledgements}
I am very grateful to have shared this exciting and fruitful research
with Tony Thomas, and wish him all the best for his 60th --- Happy
Birthday!





\bibliographystyle{aipproc}   


\end{document}